\documentclass[twocolumn,showpacs,preprintnumbers,amsmath,amssymb]{revtex4}
\usepackage{graphicx}% Include figure files
\usepackage{dcolumn}% Align table columns on decimal point
\usepackage{pifont}
\usepackage{bm}% bold math
\usepackage{mathrsfs}
\begin{document}

%\preprint{APS/123-QED}

\title{Hollow-core infrared fiber incorporating metal-wire metamaterial}

\author{Min Yan}
\email{miyan@fotonik.dtu.dk}
\author{Niels Asger Mortensen}
\email{asger@mailaps.org}
\affiliation{%
Department of Photonics Engineering (DTU Fotonik), \\
Technical University of Denmark
DK-2800 Kgs. Lyngby, Denmark
}%
%\date{\today}% It is always \today, today,
             %  but any date may be explicitly specified
\begin{abstract}
Infrared (IR) light is considered important for short-range wireless communication, thermal sensing, spectroscopy, material processing, medical surgery, astronomy etc. However, IR light is in general much harder to transport than optical light or microwave radiation. Existing hollow-core IR waveguides usually use a layer of metallic coating on the inner wall of the waveguide. Such a metallic layer, though reflective, still absorbs guided light significantly due to its finite Ohmic loss, especially for transverse-magnetic (TM) light. In this paper, we show that metal-wire based metamaterials may serve as an efficient TM reflector, reducing propagation loss of the TM mode by two orders of magnitude. By further imposing a conventional metal cladding layer, which reflects specifically transverse-electric (TE) light, we can potentially obtain a low-loss hollow-core fiber. Simulations confirm that loss values for several low-order modes are comparable to the best results reported so far.
\end{abstract}
% 42.25.Bs 	Wave propagation, transmission and absorption
%78.66.Sq 	Composite materials

\maketitle

%%%%%%%%%%%%%%%%%%%%%%%%%%  body  %%%%%%%%%%%%%%%%%%%%%%%%%%
\section{Introduction}
Waveguides for IR light are most often inferior as compared to those for optical or microwave light. In general the lack of excellent transparent solids at IR wavelength calls for hollow-core guidance rather than solid-core guidance relying on total internal reflection (TIR). While presently the latter solution remains attractive for single-mode light guidance usually over a short distance, extension of such IR fiber to high-power light guidance is difficult due to adverse effects including Fresnel loss, thermal lensing, and low damage threshold power, etc. Without TIR, hollow-core fibers require a highly-reflective cladding mirror. At the moment, there are mainly two recognized hollow-core IR waveguiding structures, either relying on reflective metal mirrors~\cite{Harrington:IRWGReview} or a dielectric photonic band-gap (PBG) material~\cite{Temelkuran:BraggFiber} which was first explored for optical wavelengths over the past decade~\cite{Cregan:PBGFiber}. Unlike at microwave wavelength, metals at IR wavelength are less perfectly reflecting mirrors due to the presence of finite Ohmic absorption. Hence a direct IR waveguiding using a hollow metallic fiber (HMF) does not work well. The same dilemma exists for the transport of terahertz electromagnetic (EM) waves, i.e. far-IR light (refer to \cite{Bowden:TeraHertzFiber} and references therein). In practice, it is necessary to further refine the HMF by adding an additional dielectric coating on the metal cladding~\cite{Harrington:IRWGReview,Bowden:TeraHertzFiber}. Photonic band-gap guidance on the other hand remains as a potential solution, but the guiding mechanism is inherently band-gap limited and heavily relies on a highly periodic photonic crystal cladding. Bragg fibers with a 700~$\rm \mu m$-diameter core have facilitated hollow-core band-gap guidance of CO$_2$ laser radiation ($10.6{\rm \mu m}$) with a propagation loss of $\sim$1~dB/m~\cite{Temelkuran:BraggFiber}. The state-of-the-art HMF with a dielectric coating with the same core size experiences a similar loss level at this wavelength \cite{Harrington:IRWGReview}.

In this paper, we show that metamaterial holds promises for entirely new guiding mechanisms, being of neither photonic band-gap nature nor relying on the classical total-internal reflection! In fact, the potentials of especially metal-based metamaterial has already been exploited for novel hollow-core planar waveguide \cite{Schwartz:externalRef} or fiber \cite{Smith:nanolett} designs. In the present work, we propose to incorporate metal-wire metamaterial into a cladding mirror design for hollow-core IR guidance. The hollow guidance mechanism is fundamentally different from previous proposals \cite{Schwartz:externalRef,Smith:nanolett} in which a metamaterial cladding with an effective refractive index between 0 and 1 is deployed for achieving light confinement, but rather based on the fact that a highly anisotropic medium with difference signs in its effective permittivity tensor components is able to totally reflect transverse-magnetic (TM) light. TM-polarized ligth is previously considered problematic to be propagated in a hollow metallic waveguide, both in slab or fiber geometry, as it suffers significantly higher loss compared to transverse-electric (TE) light.

The paper is organized as follows. First, in Section 2 we will show why a planar interface between a metal-wire metamaterial and air can serve as a mirror for TM-polarized light at IR frequency. The condition required for such a total external reflection is stated. We quantitatively derive the angular reflectance spectrum for a silver-wire medium approximated as a homogeneous medium with a effective medium theory (EMT), proving its superior reflection performance at large incidence angles compared to a bulk silver. In Section 3, we study hollow-core IR guidance in cylindrical fibers with metal-wire medium as cladding. The fiber, which converges to the traditional HMF when the fraction of metal increases, has a drastically improved performance over HMF, as we will support by calculations based on both EMT as well as more rigorous full-vectorial finite-element simulations accounting for the mesoscopic structure of the metamaterial cladding. In Section 4, we briefly compare how our proposed metamaterial fiber differs with a HMF with a dielectric inner coating. Finally, discussion and conclusion are presented in Section 5.

\section{Planar geometry}
\subsection{Why it reflects?}
\begin{figure}[b]
\centering
\includegraphics[width=4.5cm]{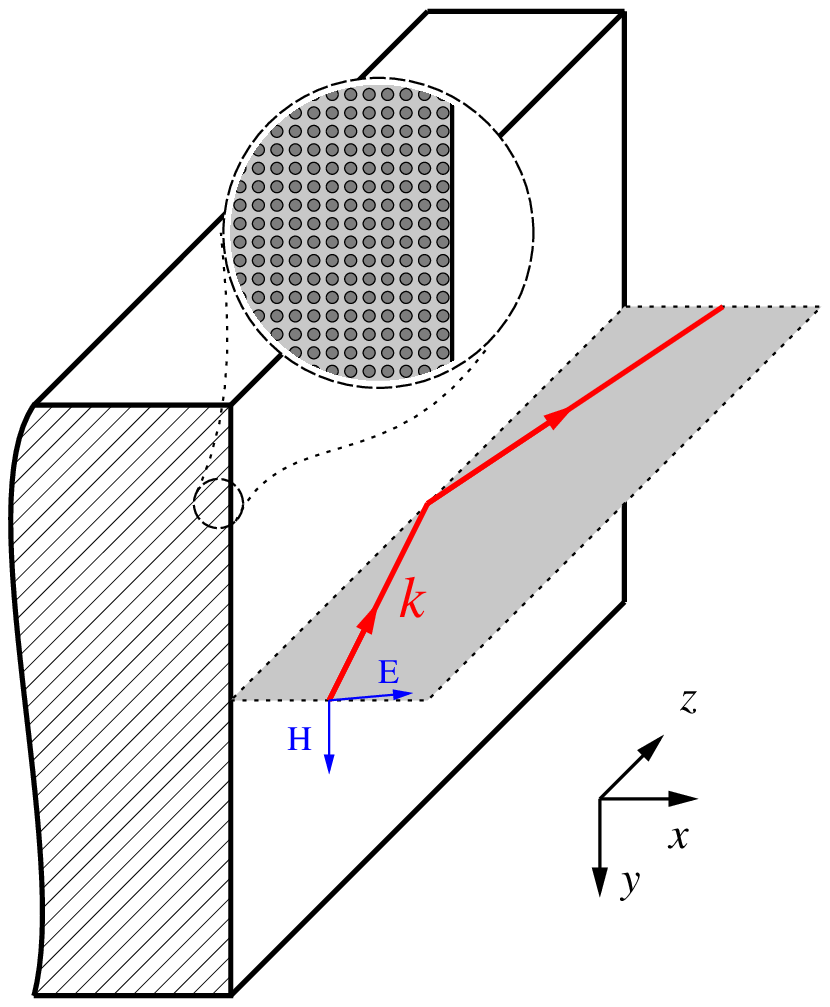}
\put(-5,-12){(a)}
\\\  \\
\includegraphics[width=8cm]{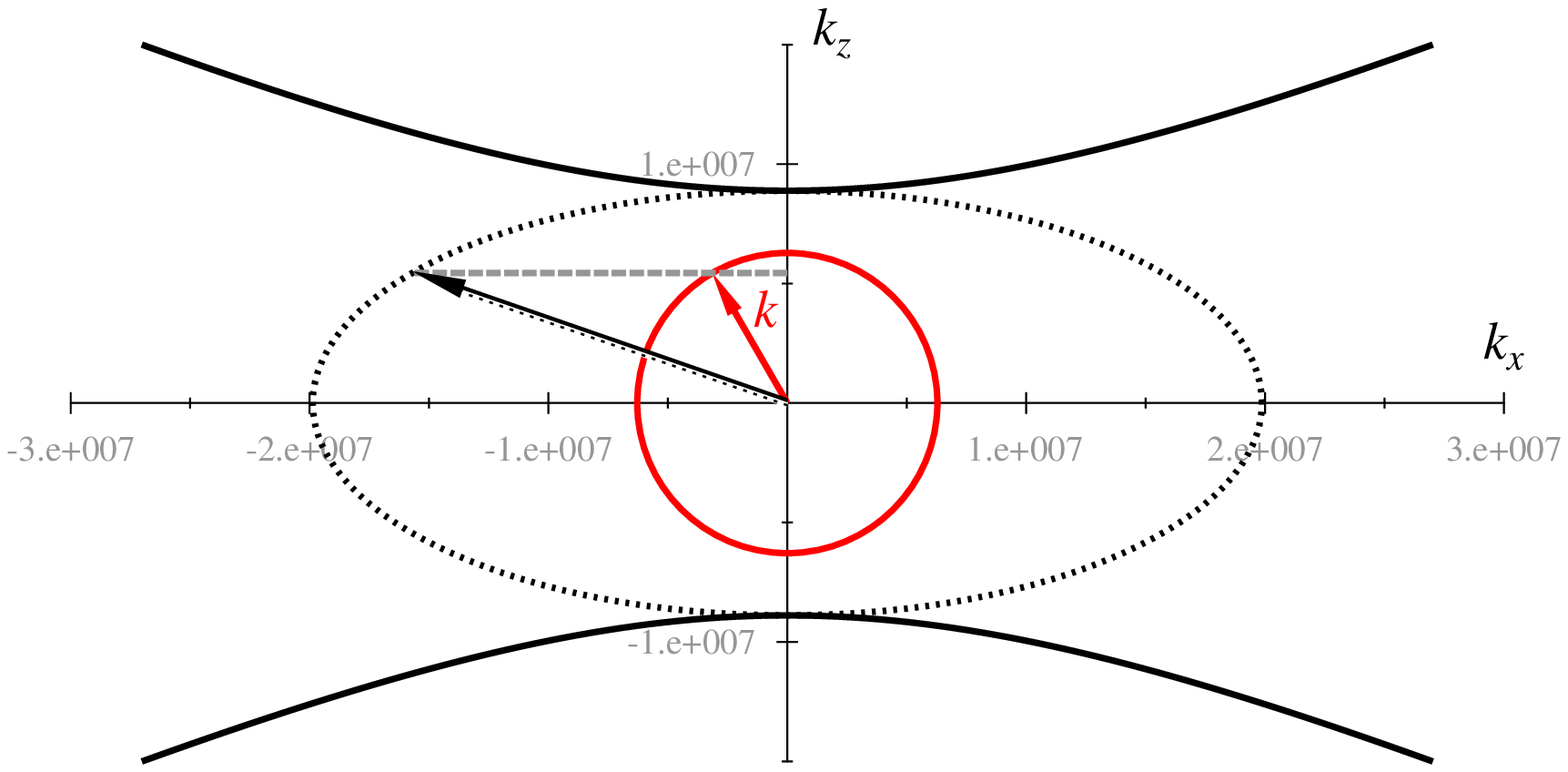}
\put(-50,12){(b)}
\caption{(a) Metal-wire medium as a TM reflector in planar geometry. (b) $k_x\sim k_z$ relations for TM light in both air (red curve) and an indefinite medium with $\epsilon_x=2, \mu_y=1, \epsilon_z=-10$ (black curves), together with wavevectors illustrating reflection for TM light incidence. Axis unit: rad/m. $\lambda=1\mu$m.}
\label{fig:slabDisp}
\end{figure}

The planar geometry serves as a simple yet clear example explaining why a metal-wire based metamaterial can easily form a reflector for TM light at IR frequency. Figure \ref{fig:slabDisp}(a) describes a semi-infinite metamaterial formed by $z$-oriented metal wires imbedded in a host dielectric occupying a half-space, say $x<0$. When the wire diameter and wire separation are much smaller than the operating wavelength, the composite can be conveniently treated as a homogenized medium with an effective permittivity tensor of the diagonal form $\overline{\overline{\varepsilon}}=\mathrm{diag}({\varepsilon_x,\varepsilon_y,\varepsilon_z})$ while for the permeability the medium appears isotropic with $\mu=1$. For uniformly dispersed wires we have $\varepsilon_x=\varepsilon_y \equiv\varepsilon_t$. When the size of metal wires is close to the skin depth of the metal (tens of nanometers), the Maxwell--Garnett theory (MGT) is adequate, especially at small metal filling fraction, for deriving the effective permittivity tensor of the homogenized metamaterial~\cite{elser:261102,Liu:08:nanowire}, i.e.
\begin{subequations}
\begin{eqnarray}
\varepsilon_t&=&\varepsilon_d+\frac{f_m\varepsilon_d(\varepsilon_m-\varepsilon_d)}{\varepsilon_d+0.5f_d(\varepsilon_m-\varepsilon_d)}  \\ \varepsilon_z&=&f_m\varepsilon_m+f_d\varepsilon_d,
\end{eqnarray}
\label{eq:MGT}
\end{subequations}
where $f_d$ and $f_m$ represent filling ratios of the dielectric and metallic materials ($f_d+f_m=1$), respectively;  $\epsilon_d$ and $\epsilon_m$ are the permittivity values of the element dielectric and metallic materials, respectively.
As shown by Eq.~(\ref{eq:MGT}), the effective permittivity components are solely determined by $f_m$. The component $\varepsilon_z$ is usually negative due to un-restricted electron motion in $z$ direction, while $\varepsilon_t$ is usually positive~\cite{elser:261102,Liu:08:nanowire,PhysRevLett.76.4773,Yao:2008:nanowire}. Such a metamaterial with both positive and negative material tensor components is referred to as an \emph{indefinite medium}~\cite{Smith:03:indefiniteMedium}.

In Fig. \ref{fig:slabDisp}(a), we schematically show that a metal-wire medium totally reflects a TM-polarized light (with field components E$_x$, H$_y$, and E$_z$) incident from air, with its wavevector lying in $xz$ plane. Before setting out to investigate the reflection conditions, we show how the total reflection is possible with a rather common indefinite medium fulfilling the very first requirement
\begin{equation}
\varepsilon_x>0,\ \mu_y>0,\ \varepsilon_z<0.   \ \ \ \ \ \mathrm{(requirement\ 1)}
\label{eq:req1}
\end{equation}
The medium chosen has $\epsilon_x=2, \mu_y=1, \epsilon_z=-10$ (a typical effective medium realizable with metal wires). Notice that only three material parameters are relevant to TM-polarized light.

When a TM wave is propagating in $xz$ plane in a general medium, the dispersion relation of the wave is governed by
\begin{equation}
\frac{k_z^2}{\varepsilon_x}+\frac{k_x^2}{\varepsilon_z}=k_0^2\mu_y,
\label{eq:slabDisp}
\end{equation}
where $k_0=\omega/c$ is the free-space wave number. In a waveguide picture where the EM field is propagating along the $z$ direction, $k_z$ is also called propagation constant $\beta$ which is required to be real. For EM wave propagating in an ordinary isotropic dielectric medium, the eligible real $(k_z,k_x)$ combinations form a circle on the $k_z$-$k_x$ plane, as shown in Fig. \ref{fig:slabDisp}(b) for the case of vacuum. However, for an indefinite medium defined by Eq.~(\ref{eq:req1}), Eq.~(\ref{eq:slabDisp}) becomes a hyperbolic equation with real $k_x$ if $k_z>k_c$, where $k_c=\sqrt{\varepsilon_x(k_0^2\mu_y-k_x^2/\varepsilon_z)}$, or otherwise an elliptic equation with imaginary $k_x$. For a waveguiding purpose, we always desire an imaginary $k_x$ in order for the field to be evanescent along the lateral $x$ direction (a wave propagating along $z$ while evanescent along $x$ is therefore a surface wave). Figure \ref{fig:slabDisp}(b) shows the $k_x\sim k_z$ dispersion relation for the above-mentioned representative indefinite medium. As indicated by Fig. \ref{fig:slabDisp}(b), the incident light from air (red arrow) can easily be matched in tangential $k$ component ($k_z$) by a surface wave in the indefinite medium (solid-dotted arrow). Therefore, the medium reflects TM light with $k$ lying in $xz$ plane. It follows that a hollow-core slab waveguide with two claddings made of such media is able to propagate TM light.

Based on the dispersion diagram in Fig. \ref{fig:slabDisp}(b), we address a further requirement in order for the total reflection to happen \emph{at all incidence angles} in $xz$ incidence plane. That is, the hyperbola shouldn't touch the red circle, otherwise at large incident angles, $k_z$ component of the incident light can be matched by a propagating wave in the indefinite medium. That is, light would transmit through the substrate. This imposes $k_0^2\varepsilon_x\mu_y>k_0^2$, or simply
\begin{equation}
\varepsilon_x\mu_y>1,   \ \ \ \ \  \mathrm{(requirement\ 2)}
\label{eq:req2}
\end{equation}
which should be imposed on top of requirement 1.

Putting the two requirements together, we obtain a relaxed (but sufficient) condition for a medium possessing all-angle reflection (in $xz$ incidence plane) for TM light, as
\begin{equation}
\varepsilon_x>1,\ \mu_y=1,\ \varepsilon_z<0.
\label{eq:slabCond}
\end{equation}
This condition is rather easy to be fulfilled with a metal-wire medium. For instance, based on MGT approximation, silver wires in a  dielectric host with any permittivity from 2 to 12, and with any metal filling fraction from 10\% to 50\%, will be able to satisfy the above condition from wavelength at $2\mu$m and beyond. This suggests the broad structural and spectral applicability of our proposed reflector and in turn waveguide designs.

\subsection{How well it reflects?}

\begin{figure}
\centering
\includegraphics[width=6.5cm]{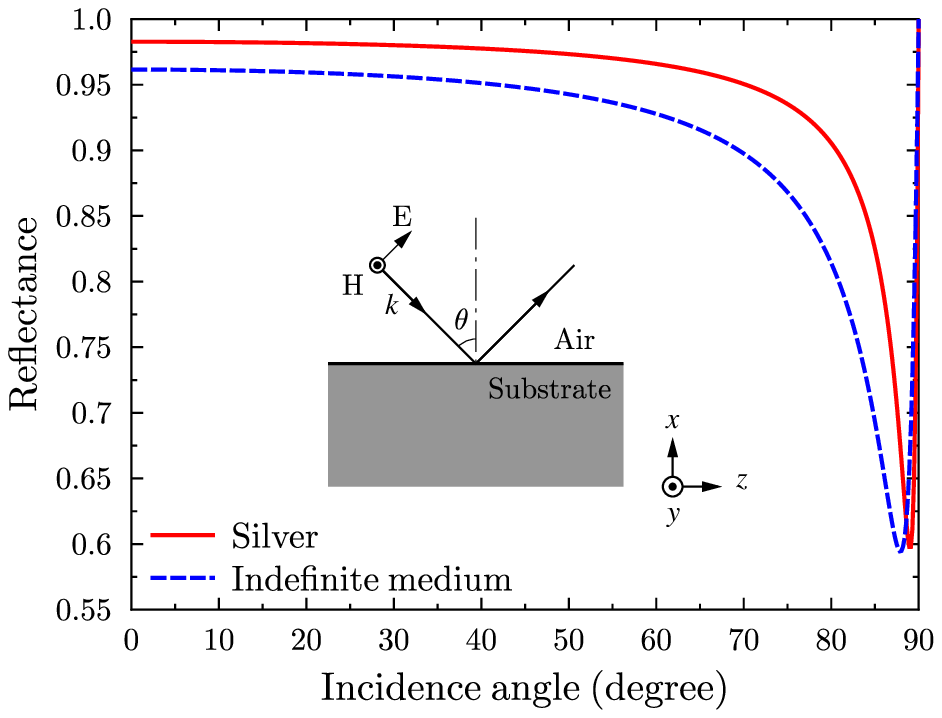}
\put(-85,-12){(a)}
\\
\includegraphics[width=6.5cm]{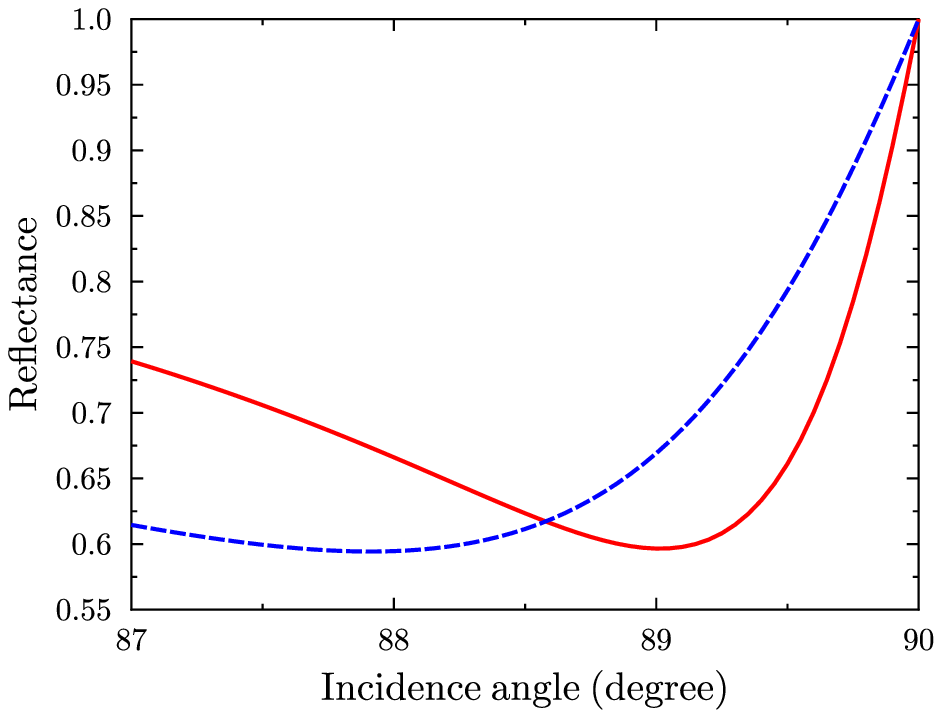}
\put(-85,-12){(b)}
\caption{Reflectance of TM-polarized incident light from two types of substrates, one silver and the other an indefinite medium derived from silver-wire-in-dielectric composite.  Wavelength is 10.6$\mu$m. (a) Reflectance spectrum for $0\sim 90$ degrees; (b) Zoom-in plot for $87\sim 90$ degrees.}
\label{fig:RefTM}
\end{figure}

Having understood the all-angle reflection for $k$ lying in $xz$ plane, our next immediate task is to know how well the reflection is. Using a full-wave analytical technique \cite{Hecht:Optics}, we examine how the reflectance of such a metamaterial substrate is compared to that of a plain metal for TM light. In this paper we concentrate primarily on the CO$_2$ laser wavelength, i.e. 10.6~${\rm\mu m}$. This wavelength is important especially due to the fact that high-power CO$_2$ laser beam can be used for material processing and various medical surgeries. For the wires we consider the use of silver with $\varepsilon=-2951+1654i$~\cite{Palik:OpticalConstMetal}, imbedded in a host dielectric with a refractive index of 2.5 corresponding to $\varepsilon = 6.25$. Several transparent materials at this wavelength have an index around this value, such as zinc selenide (ZnSe) and arsenic selenide (As$_2$Se$_3$). We consider a metamaterial in which silver wires take up a volumetric fraction of 20\%.

By Eq.~(\ref{eq:MGT}), the metamaterial under consideration can be effectively treated as a homogeneous medium with $\varepsilon_t=9.3876+0.0071i$ and $\varepsilon_z=-585.2+330.8i$, i.e. an indefinite medium. Using this approximation, the reflectance by the metamaterial-air interface as a function of incidence angle is derived in Fig. \ref{fig:RefTM}(a). The wavevector of the incoming plane wave is in $xz$ plane. The same reflectance curve is derived for a plain silver metal. Both reflectance curves are characterized by a \emph{principal angle-of-incidence} \cite{Hecht:Optics}, where the reflectance is at minimum. By comparison, it is noticed that the reflectance of the indefinite medium is lower than that of a plain metal for most incidence angles. However, as clearly shown by the zoom-in plot in Fig.~\ref{fig:RefTM}(b), at large incidence angles (greater than 88.6 degree) the metamaterial reflects better. In a large-core hollow waveguide, modes in the lowest orders indeed correspond to near-90 degree incidence angles, as will be confirmed in the next section. Therefore, a metal-wire medium can be potentially used for making a less lossy hollow waveguide as opposed to using plain metal.

\begin{figure}
\centering
\includegraphics[width=6.5cm]{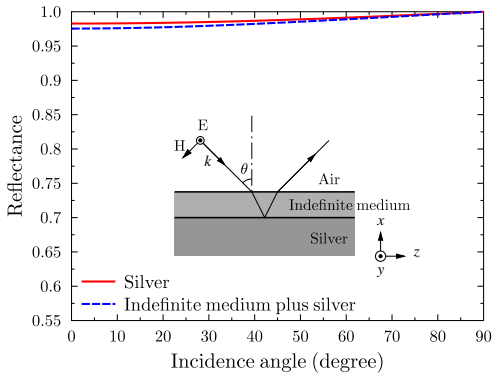}
\caption{Reflectance of TE-polarized incident light from two types of substrates, one plain silver and the other silver but with an indefinite medium layer on top (inset).  Wavelength is 10.6$\mu$m.}
\label{fig:RefTE}
\end{figure}

However, the metal-wire medium has difficulty in confining TE light, since the metamaterial appears to TE light as if it is a normal dielectric. To remedy the problem, we further add a metal substrate beneath the metamaterial. In Fig. \ref{fig:RefTE}, we show reflectance from such a hybrid substrate. The metamaterial layer is the same as that we considered in Fig. \ref{fig:RefTM} and has a thickness of 2$\mu$m. The reflectance spectrum is compared to that from a plain metal, also shown in Fig. \ref{fig:RefTE}. It is observed that by adding a layer of metamaterial the reflectance decreases only very slightly, with reflectance still higher than 99.89\% when incident angle is larger than $87^\circ$. From Figs. \ref{fig:RefTM} and \ref{fig:RefTE}, one also sees that a plain silver surface reflects TE light much better than TM light, especially at large incidence angles.

\section{Fiber geometry}
\subsection{Confinement condition and the proposal}
\begin{figure}
\centering
\includegraphics[width=6cm]{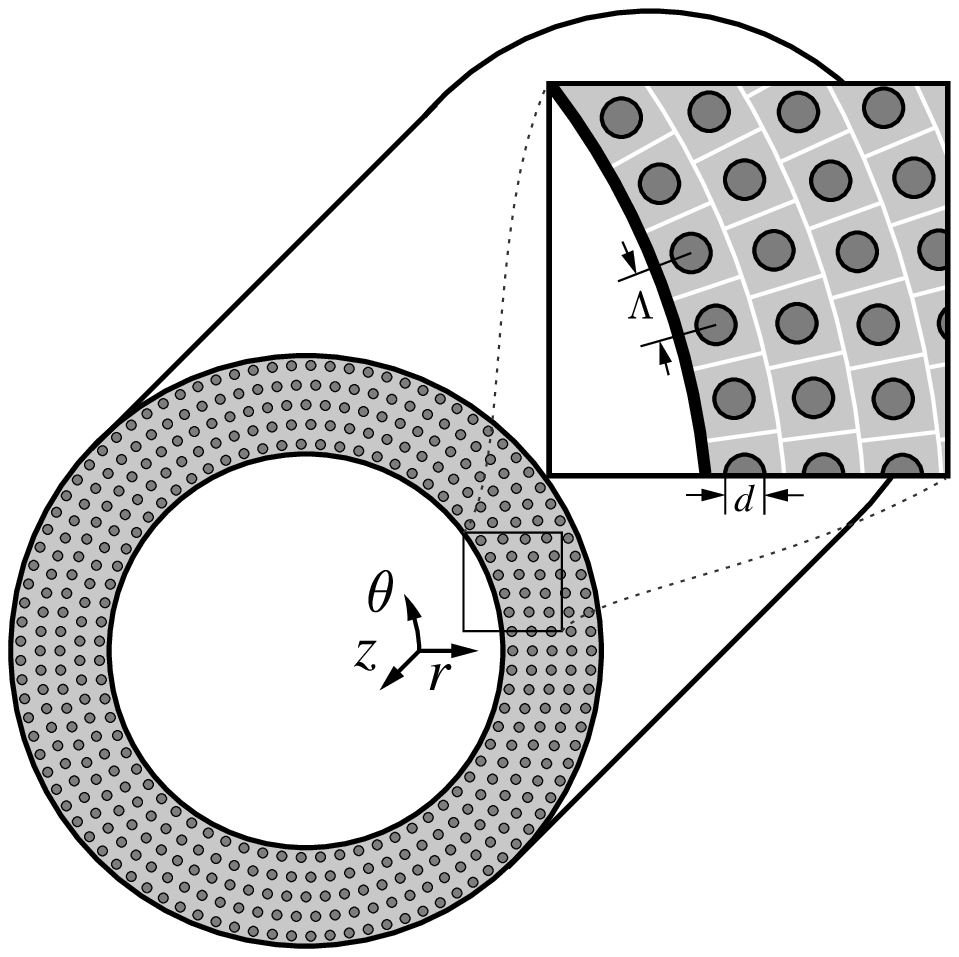}
\put(-120, -15){(a)}
\\
\includegraphics[width=6cm]{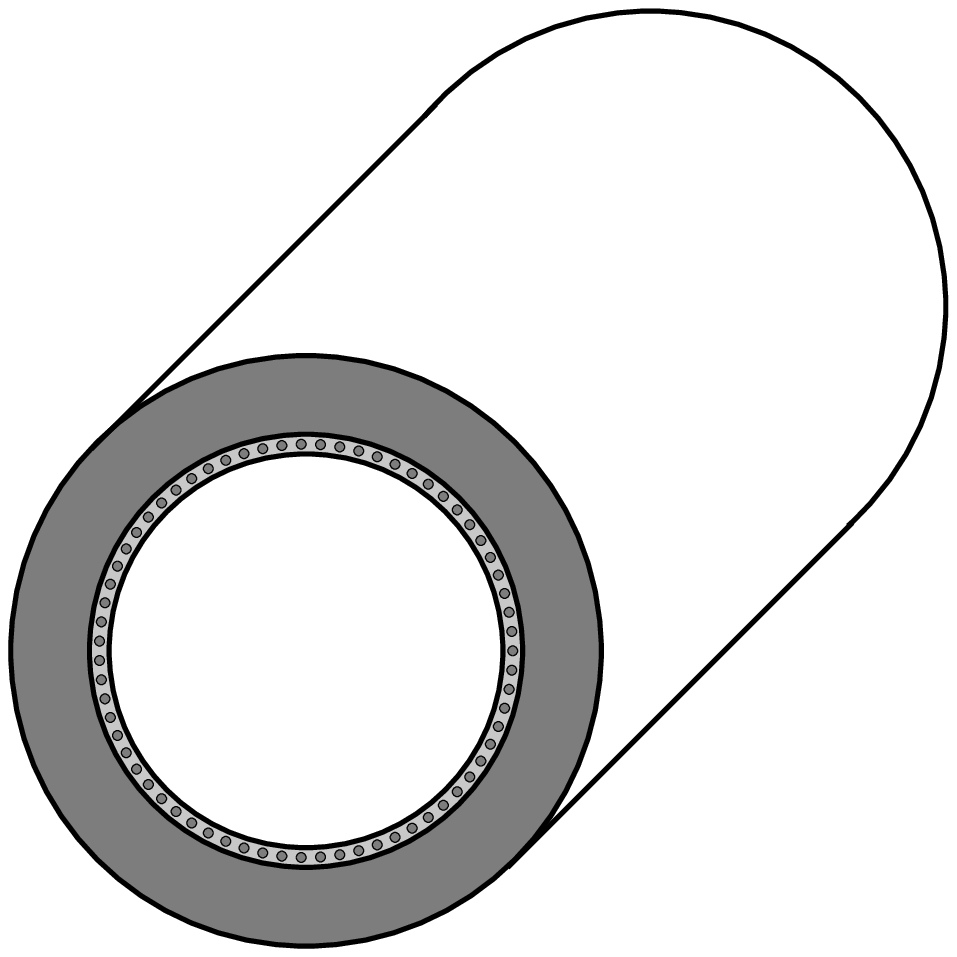}
\put(-120, -15){(b)}
\caption{Schematic diagrams for the proposed fiber structures. (a) A hollow-core fiber with a metal-wire based metamaterial cladding. Thin white lines in the inset indicate unit cells forming the metamaterial. (b) A \emph{hybrid-clad fiber}: A hollow-core fiber with a thin layer of metamaterial as its inner cladding, and a bulk metal as its outer cladding. Dark grey regions are metal. Light grey region denotes dielectric material.}
\label{fig:fiberSchematic}
\end{figure}

It is relatively straightforward to understand that the metal-wire medium discussed in Section 2 can be rolled into a fiber geometry, with the metal wires running along the fiber axis, to propagate TM light in a hollow core [see Fig. \ref{fig:fiberSchematic}(a)]. Now the TM light, in the fiber's native cylindrical coordinate, consists of E$_r$, H$_\theta$, and E$_z$ field components. Here to be more rigorous, we re-deduce the material requirement for confinement in cylindrical coordinate system.

For a fiber with an air core, one can derive the most general requirement for achieving TM field confinement (see appendix), expressed in terms of the cladding material parameters, as
\begin{equation}
\frac{\varepsilon_z}{\varepsilon_t}(k_0^2\mu_t\varepsilon_t-\beta^2)<0.
\label{eq:conditionTM}
\end{equation}
Here, $\beta$ is the propagation constant of confined mode while $\mu_t$ and $\varepsilon_t$ are cladding material parameters defined as $\varepsilon_r=\varepsilon_\theta\equiv\varepsilon_t$ and $\mu_r=\mu_\theta\equiv\mu_t$.
Similarly, the requirement for TE (with field components H$_r$, E$_\theta$, and H$_z$) confinement is
\begin{equation}
\frac{\mu_z}{\mu_t}(k_0^2\varepsilon_t\mu_t-\beta^2)<0.
\label{eq:conditionTE}
\end{equation}

Note that a hollow-core guided mode should have $\beta<k_0$.
Apparently there are several combinations of $\varepsilon_t$, $\varepsilon_z$, $\mu_t$ and $\mu_z$ which fulfil the inequalities in Eqs.~(\ref{eq:conditionTM}) and (\ref{eq:conditionTE}). Bulk metal (i.e. $\varepsilon<0$ and $\mu=1$) is of course a solution. In relating to the metal-wire metamaterial, we identify another set of solutions specifically for TM confinement, which is
\begin{equation}
\varepsilon_t>1,\ \mu_t=1,\ \varepsilon_z<0.
\label{eq:fiberCond}
\end{equation}
Notice Eq.~(\ref{eq:fiberCond}) agrees perfectly with Eq.~(\ref{eq:slabCond}). A similar solution exists for TE confinement, namely $\mu_t>1$, $\varepsilon_t=1$ and $\mu_z<0$. It is so far difficult to realize low-loss metamaterial fulfilling this TE confinement condition. In addition, plain metal is already an excellent TE reflector, as concluded in Subsection 2.2. As will be shown later, the propagation loss of the TE$_{01}$ mode can easily be 1000 times smaller than that of the TM$_{01}$ mode in a HMF. Therefore, another choice of TE reflector at a higher expense is unworthy.

The fiber structure we are to propose takes advantages of both a metal-wire medium for reflection of TM light and a plain metal for reflection of TE light. A schematic diagram of the fiber is shown in Fig.~\ref{fig:fiberSchematic}(b). The fiber cladding consists of a thin layer of metamaterial for reflecting TM light and another plain metal layer for reflecting TE light. Notice that TE light is perturbed only very slightly by the presence of the metamaterial layer, as implied by Fig.~\ref{fig:RefTE}. We refer to this fiber as a \emph{hybrid-clad fiber}. For practical realizations, the fiber structure may be coated with an outer jacket for better mechanical stability and ease of handling.

\subsection{Numerical results}
In our numerical characterization, we first turn attention to the fiber illustrated in Fig. \ref{fig:fiberSchematic}(a), whose cladding is all metamaterial. The metal wires (diameter $d$) in the metamaterial are arranged in annular layers, supported by a dielectric host. In our study cases, the metamaterial can be approximately considered as a stack of square cells [inset in Fig. \ref{fig:fiberSchematic}(a)] with cell separation $\Lambda$. For such stacking, the filling fraction of metal wires can reach up to $\sim 0.785$.

In a HMF with a relatively large core size, there are normally a huge number of propagating modes. Among them, TE$_{01}$ mode is recognized as the least lossy mode in a HMF (without dielectric coating) \cite{Bowden:TeraHertzFiber}; the first mixed-polarization MP$_{11}$ mode, or traditionally known as the HE$_{11}$ mode, is most useful for power delivering applications, due to its close resemblance to a laser output beam both in intensity profile and polarization; modes with TM polarization are the most lossy ones. Mixed-polarization (MP) modes have both TE and TM polarization components, therefore a MP mode usually has a propagation loss larger than that of a TE mode but smaller than that of a TM mode when the modes under comparison belong to the same mode group (e.g. MP$_{21}$, TE$_{01}$, and TM$_{01}$). In this paper it is of high priority to address the improved performance of TM modes especially TM$_{01}$ mode with our metamaterial fiber design. Generally, a fiber with a less lossy TM$_{01}$ mode will also have a less lossy MP$_{11}$ mode, as will be shown by our numerical calculations. In addition, the TM$_{01}$ mode is in fact valuable for other types of uses on its own. The mode has an annular beam shape and a radially-oriented electric field, which allows it to be focused into a tighter spot (compared to a linearly polarized beam) with a large longitudinal electric field component at the focus. This can be exploited for imaging or material processing, etc~\cite{Quabis:tighterFocus}.

\begin{figure}
\centering
\includegraphics[height=7.cm]{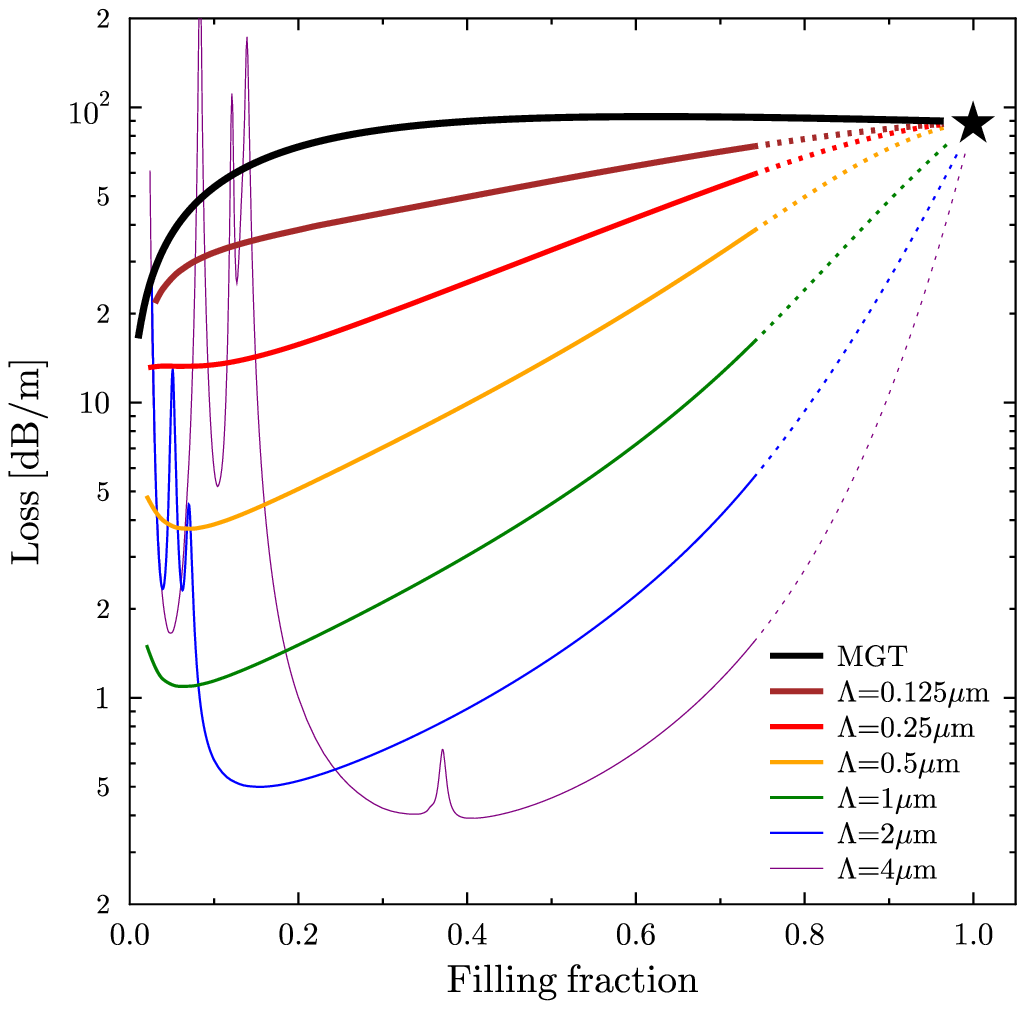}
\put(-95,-12){(a)}
\\
\includegraphics[height=7.cm]{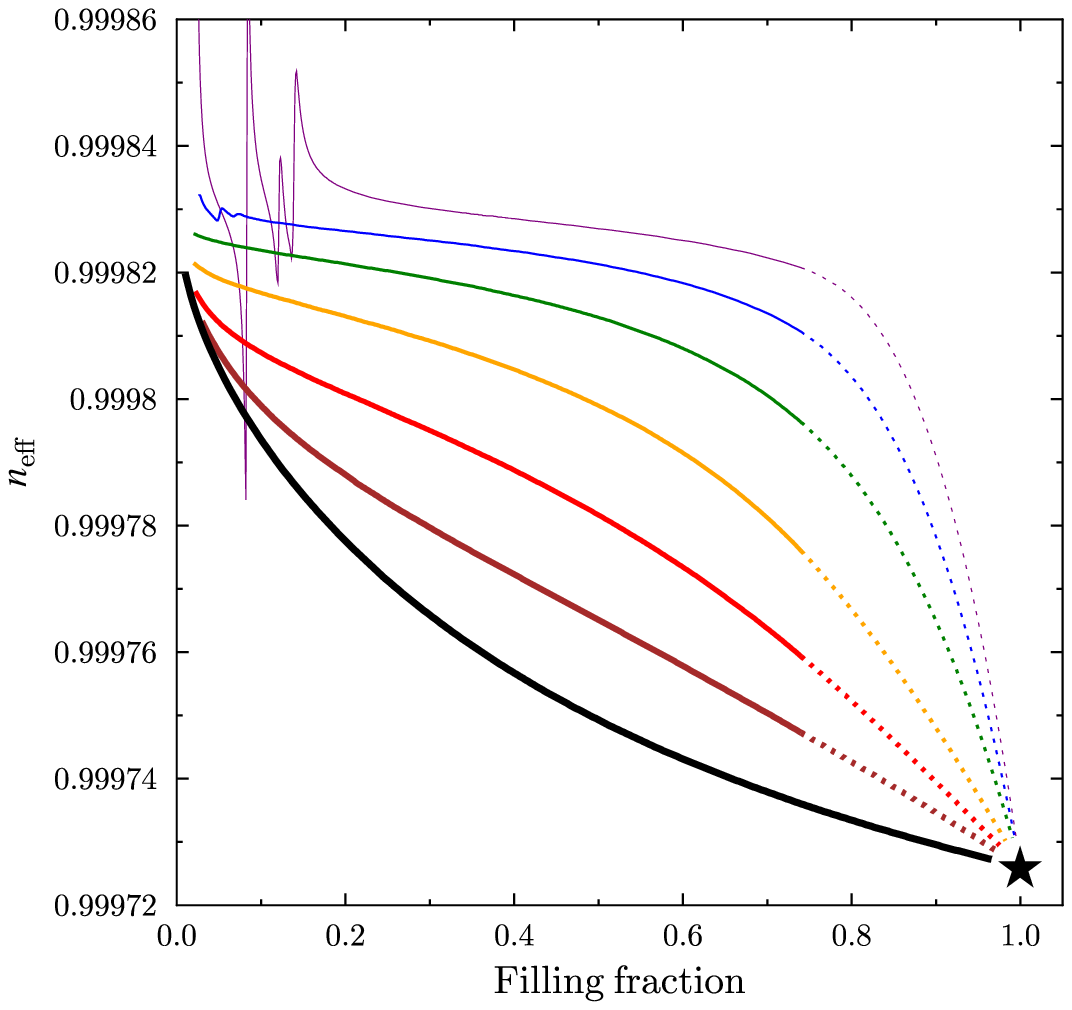}
\put(-95,-12){(b)}
\caption{(a) Propagation loss, and (b) effective mode index of the TM$_{01}$ mode as a function of metal filling fraction. The values for the mode when cladding is made of full metal are marked as \ding{72}.}
\label{fig:effect_f}
\end{figure}

We consider a fiber with all-metamaterial cladding and a core with diameter of 700~$\mu$m. With a homogenized material based on MGT approximation, we are able to quite efficiently derive the guided TM$_{01}$ mode properties, including its loss and effective mode index ($n_\mathrm{eff}$). The mode properties are plotted against $f_m$ varying from $\sim$0 (completely dielectric) to 1 (completely metal) in Fig.~\ref{fig:effect_f}. It should be remarked that the accuracy of MGT might be questionable at large $f_m$ values. Nevertheless, the results based on MGT agree well with results calculated by rigorous numerical simulations, as will be shown shortly. It is seen from Fig. \ref{fig:effect_f}(a) that, as $f_m$ decreases, the loss value initially remains close to that for the metal-clad fiber, and it drops sharply when $f_m$ becomes less than 0.2. It suggests that it is possible for the metamaterial cladding to perform better than a full-metal cladding for confining the TM mode.

In practice the limit where MGT is valid, i.e. wire diameter comparable to skin depth, is rather difficult to achieve. Here we numerically simulate realistic fiber structures by taking all mesoscopic geometrical features into account. A finite element method (FEM) has been employed for this purpose. Simulations for a number of realistic structures are summarized in Fig.~\ref{fig:effect_f}, where we show the loss and $n_\mathrm{eff}$ values of the TM$_{01}$ mode as a function of $f_m$ when wire spacing $\Lambda$ takes values of 0.125, 0.25, 0.5, 1, 2, and 4~${\rm \mu m}$. Notice that FEM simulations for all curves corresponding to realistic microstructured fibers start from $f_m=0.02$ and stop at $f_m=0.74$. Further beyond that filling fraction, the values (dotted portions of the curves) are extrapolated according to simulated data. It is observed that, when $\Lambda$ gets larger, the loss curve shifts away from the limiting MGT curve, downwards to smaller values. This further evidences that the metal-wire based metamaterial makes a superior reflector for TM light compared to plain metal. However, we notice that the downward shift in loss curve is not without limit. In particular, when $\Lambda$ increases to a certain value (after 2$\mu$m in this case), the spacings between two neighboring layers of metal wires become large enough to support resonant modes. These modes experience relatively high loss due to their proximity to metal wires. The coupling from the TM$_{01}$ core mode to these cladding resonances will result in higher loss to the TM$_{01}$ mode, which are manifested by the loss spikes observed in Fig.~\ref{fig:effect_f}(a). These resonances will become especially severe for a large $\Lambda$ value. From Fig.~\ref{fig:effect_f}, it is concluded that one should take a compromise between the propagation loss and the number of cladding resonances in choosing the right $\Lambda$ (and thereafter $d$). We should mention that it is difficult for the guided core mode to be resonant with surface plasmon polaritons (SPPs) supported by individual metal wires, as the SPP guided by a single wire has $n_\mathrm{eff}>2.5$ which is significantly larger than $n_\mathrm{eff}$ of the core mode. From Fig. \ref{fig:effect_f}(b), it is shown that the $n_{\mathrm{eff}}$ value is larger at smaller $f_m$ values, generally around 0.99982. The corresponding incidence angle for guided light can be estimated, through $\theta=\sin^{-1}(\beta/k_0)=\sin^{-1}(n_{\mathrm{eff}})$, to be $\sim88.9^\circ$. This is consistent with the reflection spectrum shown in Fig. \ref{fig:RefTM}, in which at the particular incidence angle an improved reflection by a metal-wire metamaterial is predicted.

\begin{figure}[b]
\centering
\includegraphics[width=6cm]{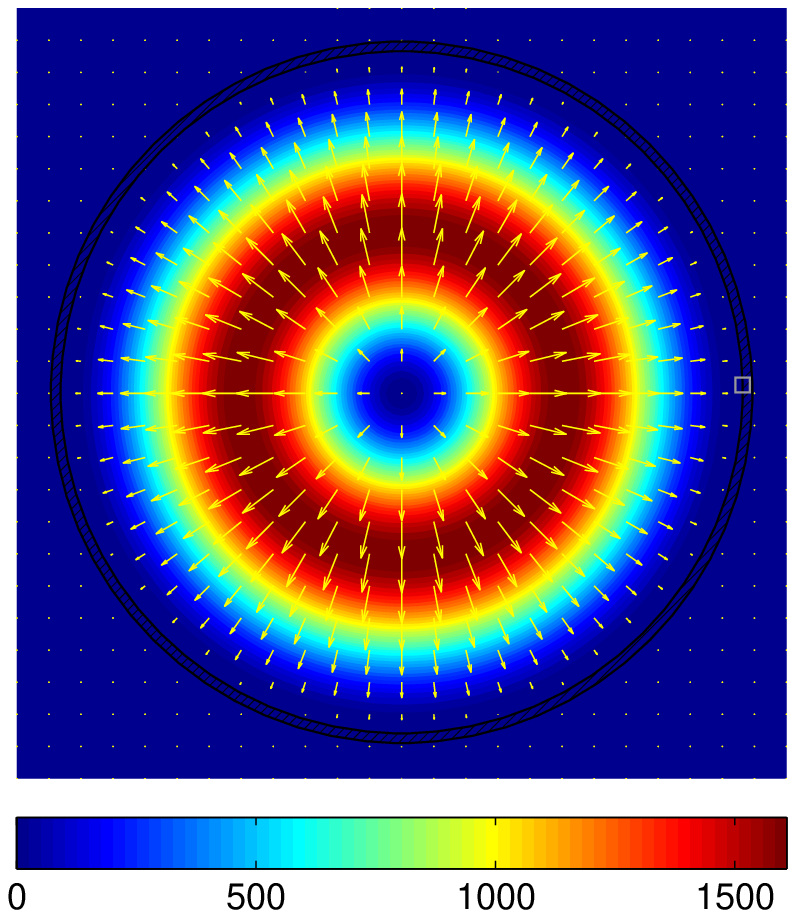}
\put(-90, -8){(a)}
\\\ \\ \ \\
\includegraphics[width=6cm]{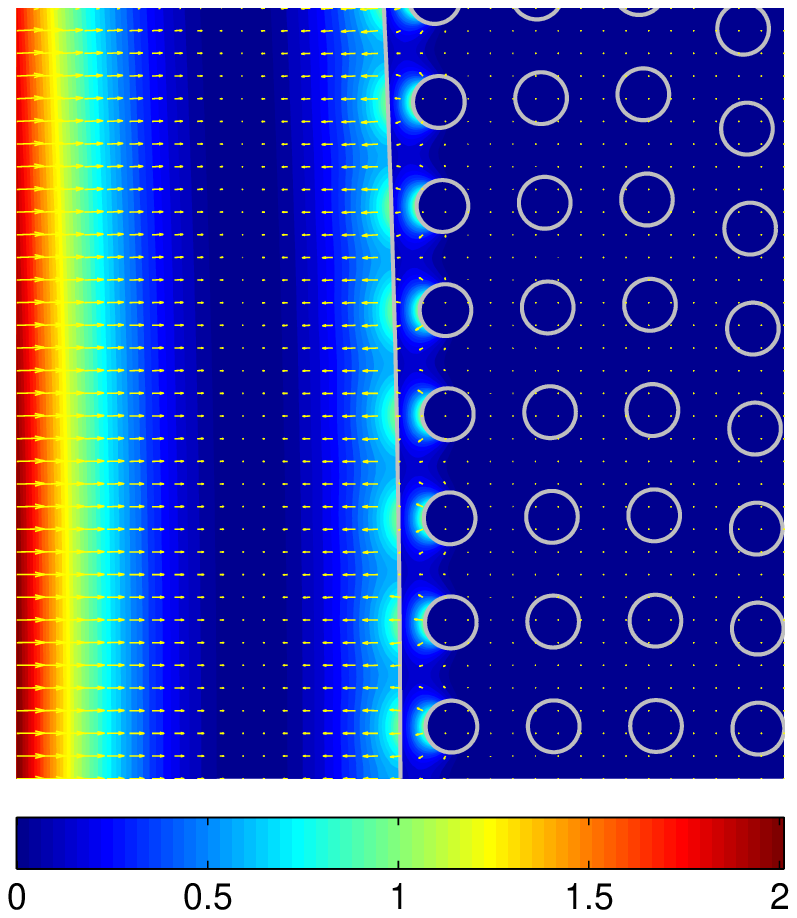}
\put(-90, -12){(b)}
\caption{Cross-sectional field distribution of the guided TM$_{01}$ mode in a metamaterial-clad fiber with a 700$\mu$m core diameter. (a) Overall mode; (b) The zoom-in plot of the region outlined in (a) by the gray line ($15\times 15\mu$m$^2$). Color shading is for the real part of the axial Poynting vector, while arrows are for transverse electric field. $\lambda=10.6\mu$m, $\Lambda=2\mu$m, $f_m=0.2$.}
\label{fig:metafiber_fields}
\end{figure}

Here we remark that, for all microstructured metamaterial fiber simulations (excluding the homogenized fiber case in MGT limiting form) in Fig. \ref{fig:effect_f}, we used five layers of metal wires in the cladding, beyond which we have air background. When $\Lambda=0.125\mu$m, the metamaterial cladding is as thin as 0.625$\mu$m (16 times smaller than 10.6$\mu$m). Quite remarkably, such a metamaterial cladding, though very thin, confines TM light exceptionally well. In Fig.~\ref{fig:metafiber_fields}, we show the field distribution of the guided TM$_{01}$ mode. It is noticed from Fig.~\ref{fig:metafiber_fields}(a) that the overall mode field is well guided by the metamaterial cladding. The zoom-in plot in Fig.~\ref{fig:metafiber_fields}(b) reveals detailed field interaction with the metal-wire medium. The \emph{partially excited} SPPs at metal wires adjacent to the interface imply the anti-resonant nature of the cladding metamaterial. The cladding field macroscopically resembles an evanescent wave which rapidly decays away from the core. The plot also quantitatively indicates that a couple of metal-wire layers would be sufficient to prevent leakage of TM light. In other words, the metamaterial cladding behaves like a metal to TM light, and it has a very small (sub-micron) \emph{effective skin depth}.

\begin{figure}
\centering
\includegraphics[width=8cm]{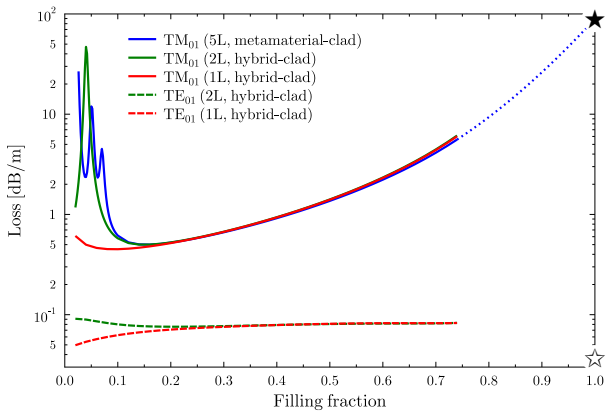}
\caption{Propagation losses of both TM$_{01}$ and TE$_{01}$ modes as a function of metal filling fraction. $\lambda=10.6\mu$m, $\Lambda=2\mu$m. The propagation loss values for the TM$_{01}$ and TE$_{01}$ modes at the $f_m=1$ limit are marked as \ding{72} and \ding{73}, respectively.}
\label{fig:hybridLoss}
\end{figure}

The fiber in Fig. \ref{fig:fiberSchematic}(a), though supporting TM modes, does neither confine TE modes nor MP modes. The hybrid-clad fiber with two claddings in Fig. \ref{fig:fiberSchematic}(b) remedies the problem. Among the two claddings, the inner one is of special importance. Based on our results in Fig.~\ref{fig:effect_f}, we here focus on a particular metal-wire spacing $\Lambda=2~{\rm \mu m}$. First we investigate the effect of the metamaterial cladding thickness by studying two hybrid-clad fibers: one has a metamaterial thickness of 2$\mu$m (one layer of metal wires); the other has a metamaterial thickness of 4$\mu$m (two layers of metal wires). The loss values for the TE$_{01}$ and TM$_{01}$ modes guided by the two fibers are shown in Fig.~\ref{fig:hybridLoss}, expressed again as functions of $f_m$. We also superposed the loss curve of the TM$_{01}$ mode guided by a fiber with pure metamaterial cladding [i.e. the curve marked with ``$\Lambda=2\mu$m'' in Fig.~\ref{fig:effect_f}(a)]. According to Fig.~\ref{fig:hybridLoss}, at a relatively large $f_m$ value the TM$_{01}$ mode loss is hardly affected as compared to the value for the metamaterial-clad fiber, which is true even when the hybrid fiber has only one layer of metal wires. At small $f_m$, a thin metamaterial cladding layer even helps to reduce the cladding resonance, and therefore to reduce the loss of the TM$_{01}$ mode. It turns out that the lowest loss for the TM$_{01}$ mode is achieved by the hybrid fiber with only one layer of metal wires. In the limit of $f_m=1.0$, both hybrid fibers degenerate into a HMF, whose TM$_{01}$ and TE$_{01}$ modes have loss values as indicated by the black and open stars, respectively, in Fig. \ref{fig:hybridLoss}. Clearly we see that in such a conventional HMF, the TM mode suffers over 1000 times higher loss as compared to the TE mode. This is the key factor that restricts the usage of such conventional fiber at this operating wavelength.

By adding a metamaterial layer as an inner cladding, the TE$_{01}$ mode is found to have a slightly higher loss as compared to that in a HMF of the same core size. However, the loss value of the TE$_{01}$ mode is still below 0.1~dB/m, which is well acceptable for a wide range of applications. No cladding resonance have been found in this particular case which deteriorates the TE propagation. However, we point out that, since the inner cladding is transparent to TE wave, it can give rise to constructive wave resonance in the layer. In turn, resonances can amplify the loss caused by metal-wire absorption. This resonance condition is fulfilled when the metamaterial thickness is equal to multiple of half transverse-wavelength for TE light in the medium, which is confirmed numerically (not shown here).

\begin{figure}
\centering
\includegraphics[width=8cm]{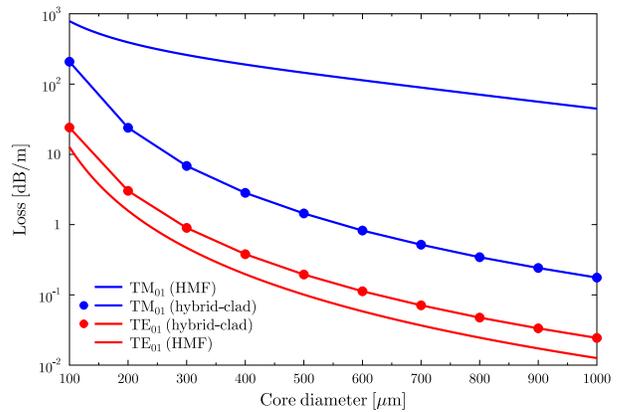}
\caption{Loss values for TE$_{01}$ and TM$_{01}$ modes guided by hybrid-clad fiber as a function of core diameter. The same curves for the full-metal fiber are also shown. $\lambda=10.6\mu$m, $\Lambda=2\mu$m, $f_m=0.2$. The inner cladding has one layer of metal wires.}
\label{fig:lossCore}
\end{figure}

In Fig. \ref{fig:lossCore}, we show how loss values of the TE$_{01}$ and TM$_{01}$ modes guided in a hybrid-clad fiber vary with respect to the fiber core size. As the core radius $R$ increases, the losses decrease roughly as $1/R^3$. Such loss dependence on $R$ is also found for other types of hollow-core waveguides, including HMF \cite{Marcatili:hollowFiber} and the OmniGuide~\cite{Johnson:01:BF}. In Fig. \ref{fig:lossCore} we also show the loss values of the same two modes guided in a HMF. By comparison, we see that propagation loss for the TM mode is greatly reduced with our hybrid-clad design (by more than 100 times when core diameter is moderately large), whereas the loss for TE mode experiences only a slight increase.

MP modes have both TE and TM field components and the two sets of field components are not independent. Through numerous simulations, we have found that a MP mode with a low azimuthal order number in general has a propagation loss in between that of the TE$_{01}$ and TM$_{01}$ modes. In addition, the loss of a MP mode is somewhat closer to that for the TE$_{01}$ mode. For example, for a fiber with a 700~${\rm \mu m}$ core diameter, the MP$_{11}$ has a loss of 0.11~dB/m, and the MP$_{21}$ mode has a loss of 0.28~dB/m. From Fig.~\ref{fig:lossCore}, losses for the TE$_{01}$ and TM$_{01}$ modes are respectively 0.07~dB/m and 0.52dB/m. It is worth mentioning that the MP$_{11}$ mode in a HMF with the same core size experiences a loss of 49~dB/m.

\section{Comparison to HMF with dielectric coating}
It has been a common practice to reduce the propagation loss of a HMF by further imposing a dielectric coating on its inner wall~\cite{Harrington:IRWGReview,Bowden:TeraHertzFiber}. Here we would like to briefly compare the performances between a dielectric-coated HMF and a metamaterial fiber. Since TE light sees the metamaterial layer as if it is a dielectric medium, the two fiber types under comparison should be more or less equivalent for guiding TE modes. The major difference lies in their guidance of TM modes. With the materials deployed in previous sections, we will show that a HMF with a dielectric coating is able to perform slightly better than a metamaterial fiber in guiding TM$_{01}$ mode. However, the former has the drawback of being quite sensitive to dielectric coating thickness, which on the other hand also implies its sensitivity to operating wavelength.

\begin{figure}[t]
\centering
\includegraphics[width=8cm]{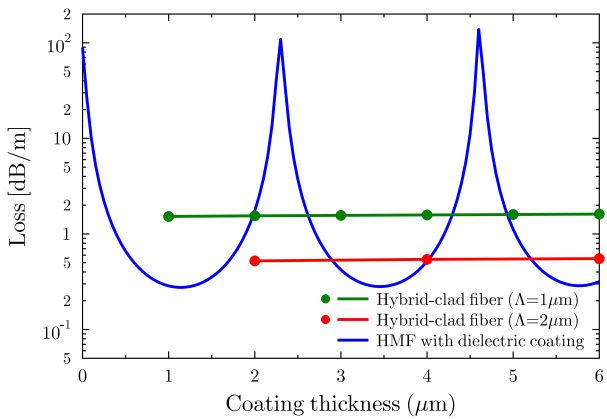}
\caption{Propagation loss for the TM$_{01}$ mode as a function of coating thickness, for a HMF with dielectric inner coating and a hollow-core hybrid fiber incorporating metamaterial.}
\label{fig:lossCompare}
\end{figure}

Let's borrow the same dielectric previously used for the metamaterial host ($\varepsilon=6.25$) as the coating for a HMF. The HMF's outer cladding (silver) starts from a fixed value $r_2=350\mu$m. The first cladding interface is located at $r_1=r_2-d_c$ with $d_c$ the coating thickness. We examine the propagation loss of the guided TM$_{01}$ mode as a function $d_c$. The result is shown in Fig. \ref{fig:lossCompare}. When the coating thickness increases from 0 to 6$\mu$m, the mode loss undergos a periodic variation, with value ranging from 0.275 to over 100$\rm dB/m$. This periodic change in loss is caused by the resonance and antiresonance of the dielectric layer. Therefore experimentally one has to locate an optimum coating thickness in order for such a fiber to work at a minimum loss~\cite{Bowden:TeraHertzFiber}. Now, if we replace the dielectric coating with a metamaterial one, the periodic variation in modal loss can be suppressed. This is shown again in Fig. \ref{fig:lossCompare} by two flat curves, which correspond to two hybrid fibers with different wire separations but with the same metal filling fraction at 20\%. The independence of the TM modal loss to metamaterial coating thickness is an inherent result of the fact that TM waves are evanescent in such metamaterial. Hence practically the hybrid fiber with metamaterial coating might be less sensitive to structural parameters, at least for the TM modes. We have also explored the possibility to further reduce the TM$_{01}$ loss of a hybrid-clad fiber with a third dielectric coating on top of the metamaterial layer. However, the further reduction in loss is not significant.

We should emphasize that the difference between the two types of fibers is far more complicated than what's been presented by Fig. \ref{fig:lossCompare}. For example unlike TM modes, TE modes in the hybrid metamaterial fiber should have a periodic loss dependence on the coating thickness. That is, the two types of fibers have similar performances for guiding TE modes. Comparison of the MP modes between two types of fibers would be more intriguing. Since a MP mode comprises both TE and TM wave components, such modes in both fibers will be sensitive to their coating thickness. Some difference should however be expected because of their different guidance behaviors for the TM wave components. Due to significantly heavier computing resources required for numerically deriving the MP modes, an explicit comparison for guidance of MP modes by the two types of fibers is not presented here.

\section{Discussion and conclusion}
Although our presentation has focused on the CO$_{2}$ wavelength, simulations were also carried out for other fiber structures based on silver and silica materials designed to operate at 1.55$\mu$m wavelength. Very similar improvement in TM mode guidance for a hybrid-clad metamaterial fiber compared to a metal-clad fiber is noticed. However, due to inherently huge Ohmic absorption of silver at this wavelength, propagation loss for the TM$_{01}$ mode in a hybrid-clad fiber, e.g. with a 30$\mu$m core diameter, though reduced, can still be higher than 100dB/m, which is hardly useful for practical applications. Hence with commonly accessible materials at disposal, the advantage of the proposed metamaterial fiber only becomes obvious at IR frequencies.

In conclusion, we have shown that a metal-wire based metamaterial functions as a TM reflector. The reflectance from a substrate made of such a metamaterial can be better than that from a plain metal at large incidence angles. Numerical simulations show that a hollow fiber with such a metamaterial cladding can easily propagate a TM mode 100 times farther compared to a simple HMF. With a hybrid-clad fiber which has a thin inner metamaterial cladding and an outer metal cladding, we can achieve low-loss propagation of modes in all categories. Such a fiber in theory can be as compelling as the state-of-the-art fibers, including the Bragg fiber and the more traditional HMF with a dielectric coating, for IR light delivery. We emphasize that the proposed guiding principle and in turn the fiber structure is within our fabrication capability at least in a foreseeable future. Our study also suggests that innovations in metamaterial technology can open many other possibilities for designing new EM waveguides, especially for radiation frequencies previously considered problematic to transport.

\section*{Acknowledgement}
This work is supported by The Danish Council for Strategic Research through the
Strategic Program for Young Researchers (DSF grant 2117-05-0037).

\section*{Appendix}
We derive the material requirement for wave confinement in a cylindrical hollow fiber as specified by Eqs.~(\ref{eq:conditionTM}) and (\ref{eq:conditionTE}) in the text. We assume that the material parameters of the cladding have the tensor form as
\begin{equation}
\overline{\overline{\varepsilon}}=
\left( \begin{array}{ccc}
\varepsilon_r & 0 & 0 \\
0 & \varepsilon_\theta & 0 \\
0 & 0 & \varepsilon_z \end{array}
\right), \ \ \
\overline{\overline{\mu}}=
\left( \begin{array}{ccc}
\mu_r & 0 & 0 \\
0 & \mu_\theta & 0 \\
0 & 0 & \mu_z \end{array}
\right).
\label{eqSI:material}
\end{equation}
Note that any tenor component can be a negative value. To simplify our analysis, we further assume $\varepsilon_r=\varepsilon_\theta\equiv\varepsilon_t$ and $\mu_r=\mu_\theta\equiv\mu_t$. The harmonic dependence is taken as $\exp(-j\omega t+\beta z)$. Due to cylindrical symmetry, field within the medium is completely characterized by two similar wave equations, one for $\mathrm{H}_z$ and the other for $\mathrm{E}_z$. The $\mathrm{H}_z$ wave equation is
 \begin{equation}
 \frac{\partial^2\mathrm{H}_z}{\partial r^2}+\frac{1}{r^2}\frac{\partial^2\mathrm{H}_z}{\partial \theta^2}+\frac{1}{r}\frac{\partial\mathrm{H}_z}{\partial r}+\frac{\mu_z}{\mu_t}k_t^2\mathrm{H}_z=0,
 \label{eq:waveMaster}
 \end{equation}
 where $k_t^2=k_0^2\mu_t\varepsilon_t-\beta^2$. By variable separation $\mathrm{H}_z=\Psi(r)\Theta(\theta)$, Eq.~(\ref{eq:waveMaster}) can be decomposed into two equations. One of them gives rise to angular dependence of the field as $\exp{(im\theta)}$, where $m$ is an integer denoting the angular momentum number. The radial dependence of the field is governed by
 \begin{equation}
 \frac{\partial^2\Psi}{\partial r^2}+\frac{1}{r}\frac{\Psi}{\partial r}+\frac{1}{r^2}\left(\frac{\mu_z}{\mu_t}k_t^2r^2-m^2\right)\Psi=0
 \label{eq:waveR}
 \end{equation}
Equation~(\ref{eq:waveR}) is a Bessel or modified Bessel differential equation, depending on the sign of $\frac{\mu_z}{\mu_t}k_t^2$. For light confinement in a hollow core, it is necessary for the field to be evanescent in the cladding while $\beta^2<k_0^2$ (as wave should be propagating in the core). Subsequently, we find that this condition can be fulfilled when
\begin{equation}
\frac{\mu_z}{\mu_t}(k_0^2\varepsilon_t\mu_t-\beta^2)<0
\label{eq:conditionHz}
\end{equation}
and the corresponding radial eigen-field in the cladding can be written generally in Bessel functions as
\begin{equation}
\Psi=\mathcal{A}I_m(\tilde{k}_tr)+\mathcal{B}K_m(\tilde{k}_tr),
\label{eq:solutionHz}
\end{equation}
where $\tilde{k_t}=\sqrt{-\frac{\mu_z}{\mu_t}k_t^2}$. Similar analysis can be carried out for the $\mathrm{E}_z$ wave equation. And the resulted condition for $\mathrm{E}_z$ confinement is
\begin{equation}
\frac{\varepsilon_z}{\varepsilon_t}(k_0^2\mu_t\varepsilon_t-\beta^2)<0.
\label{eq:conditionEz}
\end{equation}
The general radial wave solution is the same as Eq.~(\ref{eq:solutionHz}) except with $\mu$ changed to $\varepsilon$. Other field components can be written as a function of $\mathrm{E}_z$ and $\mathrm{H}_z$. Therefore once the conditions specified by Eqs.~(\ref{eq:conditionHz}) and (\ref{eq:conditionEz}) are fulfilled, confinement of the overall mode can be ensured.

When $m=0$, the six field components are divided into two unrelated groups giving rise to two types of angularly invariant modes: a TE mode group with field components H$_z$, E$_\theta$, and H$_r$; and a TM mode group with field components E$_z$, H$_\theta$, and E$_r$. Therefore, Eq.~(\ref{eq:conditionHz}) is the general guidance condition for TE modes, and Eq.~(\ref{eq:conditionEz}) is the general guidance condition for TM modes.

\end{document}